\documentclass[aps, prd, showpacs, superscriptaddress, groupedaddress, twocolumn, floatfix]{revtex4-2}

\usepackage{graphicx}% Include figure files
\usepackage{yfonts}
\usepackage{amsmath}
\usepackage{amssymb}
\usepackage{amsfonts}
\usepackage{amsthm}
\usepackage{mathtools}
\usepackage{dcolumn}% Align table columns on decimal point
\usepackage{listings}
\usepackage{subfigure, rotating, bm, array}
\usepackage[utf8]{inputenc}
\usepackage[english]{babel}
\usepackage[pagebackref=false, colorlinks=true]{hyperref}
\hypersetup{
linkcolor=blue,     % color of internal links
citecolor=blue,     % color of links to bibliography
urlcolor=blue}      % color of the url

\usepackage{float}

\begin{document}
%\preprint{APS/123-QED}
%\preprint{}
\title{Post-Newtonian properties of EMRI with Power Law Potential}%{Manuscript Title:\\with Forced Linebreak}% Force line breaks with \\
%\thanks{10th September 2020}%
\author{Chinmay N. Gandevikar}%
\email{cgandevikar@gmail.com}
\affiliation{BITS Pilani K.K. Birla Goa Campus, Sancoale, Goa 403726, India}
 %Lines break 
\author{Divyesh N. Solanki}
\email{divyeshsolanki98@gmail.com}
\affiliation{Sardar Vallabhbhai National Institute of Technology, Surat GUJ 395007,  India}
\author{Dipanjan Dey}
\email{deydipanjan7@gmail.com}
\affiliation{International Center for Cosmology, Charusat University, Gujarat 388421, India}

%\affiliation{%
% This line break forced with %\textbackslash\textbackslash
%}%


%\affiliation{
%\affiliation{
%\affiliation{%



\begin{abstract}
There are many astrophysical scenarios where extreme mass ratio inspiral (EMRI) binaries can be surrounded by matter distribution. The distribution of mass can affect the dynamical properties (e.g. orbital frequency, average energy radiation rate, etc.) of the EMRI. In this matter distribution, instead of Kepler-Newton potential, one may consider a more general form of potential i.e. power law potential. Moreover, due to the power law potential, the velocity profile of  test particles does not fall as much as that predicted by Kepler-Newton potential and this feature of the velocity profile may be observationally important. 
%The average energy radiation rate from an extreme mass ratio inspiral (EMRI) and the orbital frequency of the EMRI is very interesting as observing these is one of the goals of the upcoming gravitational wave detector-LISA.
In this study, we have obtained the first post-Newtonian (1PN)  expressions for dynamical quantities and the average energy radiation rate from the circular orbit EMRI which is surrounded by a matter distribution. We show that the energy radiation rate and orbital frequency of EMRI can be significantly different in the presence of power law potential as compared to that in the Kepler-Newton potential, signatures of which may be observed in gravitational waves from EMRI.  %In the case of Milky way the energy radiation is ($10^12$) times higher than that expected from a similar system in Kepler-Newton potential.
%\begin{description}
%\end{description}

%\bigskip
\end{abstract}

\maketitle

%\tableofcontents

\section{\label{sec:Intro}Introduction %First-level heading:\protect\\ The line
%break was forced \lowercase{via} \textbackslash\textbackslash
}
%Two bodies under mutual gravity go around each other forming a binary system. Today these systems are very important to researchers. They are the only source for detectable gravitational waves.

In a galaxy, the study of dynamics of the bodies around the central object has been of great interest for the last four decades. According to Kepler-Newton potential, the velocities of the stars about a central supermassive object in a galaxy should drop drastically with increasing radial distance. However, observations suggest otherwise. The Kepler-Newton expression (also denoted as KN potential) of gravitational potential ($\frac{Gm}{r}$) known to us can be obtained from the effective potential of the Schwarzschild spacetime at an infinite distance from the central body. However, the Schwarzschild metric is a vacuum solution of the Einstein's field equation. But at the galactic scale, due to dark matter and baryonic matter, the system is non-vacuum. Due to this limitation, the Schwarzschild metric or the KN potential cannot theoretically explain galactic dynamics. 

In the late  1970s  and early  1980s,  it was confirmed by the extensive study reported in Sofue and  Rubin in 2001 \cite{Sofue_2001} that most of the stars in a galaxy rotate at nearly the same velocities. This means that the velocities of these stars are independent of their distance from the central mass. Investigation of this novel observation has been a hot topic for several decades. There have been multiple theories like modified Newtonian dynamic (MOND) \cite{MOND}, the existence of dark matter, modified gravity, etc. \cite{evs, Rodrigues, boris}. In 1933, Zwicky \cite{zwicky} suggested ``missing mass” to account for the orbital velocities of galaxies in clusters. Later, V Rubin and their group studied the possibility of dark matter around galaxies \cite{Rubin80, Rubin70}. In \cite{Diaz, Donato}, authors study the mass density distribution in galaxies and compare them with the observations. The density profile of dark matter famously known as the Navvaro-Frenk-White (NFW) profile \cite{NFW} in the galaxy justifies the nature of the curve. Dey et al. have worked on developing a general relativistic approach to understand galactic dynamics. Bertrand Spacetimes is one of the non-vacuum spacetime which can be thought to be seeded by dark matter \cite{Dey13, Dey13B,Dey17,Dey15}. In the Newtonian limit, this spacetime can effectively give the NFW dark matter density profile and hence can explain the observed behaviour of rotational velocity curves. In 1983, Milgrom suggested modifying Newtonian dynamics (MOND) which is an alternate way to justify the curves. According to the MOND, Newton's laws of gravity get modified at a large distance scale. The theoretical prediction of MOND satisfactorily fit the observed velocity profiles of stars for a large number of galaxies \cite{lih}. Later, relativistic study of MOND was done by Bekenstein \cite{bekenstein}, Moffat \cite{moffat2005,moffat} and Brownstein \cite{Brownstein_2006}.

Logarithmic potential and appropriate power law potential can lead to the rotational velocities (matching the observations) higher than that in the case of KN potential. The velocity curve plotted using the logarithmic potential is completely flat. Recently, Munera and  Delgado-Correal \cite{Munera} have shown derivation of logarithmic potential as a solution of non-homogeneous Laplace equation.
Power law forces are mentioned in classical books like Goldstein \cite{Goldstein} and Danby \cite{Danby}. The power law potentials (also denoted as PL potential) have been extensively studied in the non-relativistic case. In \cite{Valluri_2012, Valluri_2005, Lynden-Bell}, the nature of orbits in the presence of power law potential is studied extensively.

The KN potential can be written as an approximate potential derived from the Schwarzschild solution of Einstein field equations which is only valid in vacuum region. Hence, it is not entirely applicable to study binary systems in galaxies or systems with dark matter distribution around them. Power law potential can incorporate the effect of this mass distribution and hence also can justify the behaviour of the velocity profile. In this paper, we consider an extreme mass ratio inspiral (EMRI) binary of a stellar object which is orbiting the central supermassive object in power law potential. The orbital motion makes the EMRI radiate its energy which propagates as the gravitational waves. We consider that this stellar object is in the faraway region from the central supermassive object. This region allows the usability of Post-Newtonian (PN) theory to obtain more precise expressions for the quantities that can be described for the system, e.g., orbital velocity of the object, the orbital radius, the acceleration of the object etc. Subsequently, we predict the average energy radiation rate from the system. The upcoming advanced gravitational-wave detectors like LISA may be capable of detecting the gravitational radiations from EMRI binaries \cite{Barack_2004, Amaro_2012, Glampedakis_2005, Gair_2013}.

We organize the paper in the following way. In Sec.~(\ref{two}), we derive the acceleration of a body up to 1PN approximation using power law potential instead of KN potential. The binary system and the PN corrected expressions for the dynamical quantities are discussed in Sec.~(\ref{three}). In Sec.~(\ref{four}), we discuss the PN corrected average energy radiation rate. In Sec.~(\ref{five}), the significant dynamical variables like the velocity and the orbital frequency, and energy radiation in the case of power law potential are compared with that in the case of KN potential system. In Sec.~(\ref{six}), we discuss the important results and possible future works.

%This sample document demonstrates proper use of REV\TeX~4.2 (and
%\LaTeXe) in mansucripts prepared for submission to APS
%journals. Further information can be found in the REV\TeX~4.2
%documentation included in the distribution or available at
%\url{http://journals.aps.org/revtex/}.

%When commands are referred to in this example file, they are always
%%shown with their required arguments, using normal \TeX{} format. In
%this format, \verb+#1+, \verb+#2+, etc. stand for required
%author-supplied arguments to commands. For example, in
%\verb+\section{#1}+ the \verb+#1+ stands for the title text of the
%author's section heading, and in \verb+\title{#1}+ the \verb+#1+
%stands for the title text of the paper.%

%Line breaks in section headings at all levels can be introduced using
%\textbackslash\textbackslash. A blank input line tells \TeX\ that the
%paragraph has ended. Note that top-level section headings are
%automatically uppercased. If a specific letter or word should appear in
%%lowercase instead, you must escape it using \verb+\lowercase{#1}+ as
%in the word ``via'' above.

\section{\label{sec:eom}Acceleration of the reference body in the presence of power law potential%First-level heading:\protect\\ The line
%break was forced \lowercase{via} \textbackslash\textbackslash
}
\label{two}
The first order approximate expression for acceleration vector ($a_P^j$)  of a body named as body-P can be written as \cite{Eric_Will},
\begin{eqnarray} \label{acc_eqn}
 a^j_P = \partial^j U_P + \frac{1}{c^2} [ (v_P^2-4U_P)\partial^jU_P-4v_P^jv_P^k\partial_kU_P\nonumber\\   -3v_P^j\dot{U_P} + \partial^j \psi_P     +\frac{1}{2}\partial^j \ddot{X_P}  + 4\dot{U_P}^j \nonumber\\  - 4(\partial^jU_P^k - \partial^kU_P^j)v_{P,k} ] \nonumber\\ + O(c^{-4}),
\end{eqnarray}
where $U_P$, $U^j_P$, $\psi^j_P$ and $X_P$ are the external scalar gravitational potential, vector gravitational potential, post-Newtonian correction to scalar gravitational potential and the super-potential respectively.

The components of the EMRI under consideration are a supermassive compact object-$Q$ of mass $M_Q$ and a stellar mass object-$P$ of mass $M_P$ (which is much less than that of body-$Q$). We use the following notations throughout the paper: $\mathbf{y}_Q \text{ and }\mathbf{y}_P$ are the position vectors of the bodies Q and P in the center of mass frame,
$r_{PQ}$ is the distance between the two bodies, also given as $|\mathbf{y}_P-\mathbf{y}_Q|$.
$\mathbf{v}_Q$ and $\mathbf{v}_P$ are the absolute velocities of the bodies,
$\mathbf{a}_Q$ and $\mathbf{a}_P$ are the absolute accelerations of the bodies,
$\mathbf{x}_S$ is the position vector of any field point `$S$' where potential is being evaluated.
The unit vector in $\mathbf{x}_{SQ}$ direction and the unit vector in the direction of $\mathbf{x}_{PQ}$ can be written as $\mathbf{\hat{n}}_{SQ}=\frac{\mathbf{x_{SQ}}}{|\mathbf{x_{SQ}}|}=\frac{\mathbf{x}_S-\mathbf{y}_Q}{|\mathbf{x}_S-\mathbf{y}_Q|}$ and $\mathbf{\hat{n}}_{PQ}=\frac{\mathbf{x_{PQ}}}{|\mathbf{x_{PQ}}|}=\frac{\mathbf{y}_P-\mathbf{y}_Q}{|\mathbf{y}_P-\mathbf{y}_Q|}$ respectively.

With increasing radial distance from the central supermassive object, according to Kepler's laws, the rotational velocities are expected to fall down to nearly zero. However, in order to have higher values of rotational velocities (which is generally observed in galactic scales \cite{Sofue_2001, Rubin70, Rubin80}), one can use specific forms of power law potential \cite{Struck_2006,Struck_2015}.
A general form of the expression of the Newtonian order (i.e. 0 PN order) gravitational acceleration $\mathbf{a}_{P,N}
(r)$ for the power law potential can be written as, 
\begin{equation} \label{Nacc}
    \mathbf{a}_{P,N}(r_{PQ})=\partial^jU_P (r_{PQ})=-\frac{GM^*_Q r_
    {PQ}}{(\epsilon^2+r_{PQ}^2)^{\delta +1}} \mathbf{\hat{n}}_{PQ}.
\end{equation}
Here, $M^*_Q$ is the constant scale mass of the potential due to body-Q, 
%$r_{PQ}$ is the distance from the body to the field point where the potential is being evaluated, 
$\epsilon$ is the scale radius of the source and $\delta$ can have any real value \cite{Binney}. Note that the subscript in the acceleration is ``$P,N$" denoting acceleration of body-P at Newtonian order, which is different from ``$PN$" which denotes the post-Newtonian corrections.
Two interesting special cases are
\begin{enumerate}
    %\item The rotational circular velocity is linearly dependent on radial distance ($v_{rot}\propto r$): $\delta=-1$
    \item The circular velocity is constant ($v_{circ}=\sqrt{|\mathbf{a}|r} =$ const) i.e. $\delta=0$ and
    \item The circular velocity is inversely proportional to the square root of the radial distance ($v_{circ}=\sqrt{|\mathbf{a}|r} \propto r^{-1/2}$) i.e. $\delta=\frac12$.
\end{enumerate}
Above two cases correspond to flat and Keplerian rotation curves, respectively.
The exactly flat rotation curves can be obtained at $\delta = 0$. Using Eq. (\ref{Nacc}), it can be seen that the corresponding potential is the logarithmic potential. In this study, the general power law potential is taken under consideration except for the logarithmic potential case. 
The potential ($U_P$) corresponding to the above mentioned acceleration (Eq. (\ref{Nacc})) at point P due to super-massive body-Q \cite{Struck_2006} is
\begin{equation} \label{NU}
     U_P= \frac{G M^*_Q}{2 \delta (\epsilon^2 +r_
    {PQ}^2)^\delta}.
\end{equation}
Corresponding vector potential ($U_P^j$) \cite{Eric_Will} is accordingly obtained 
\begin{equation} \label{NUj}
    U^j_P=\frac{G M^*_Q }{2 \delta (\epsilon^2 +r_
    {PQ}^2)^\delta}{v}^j_Q .
\end{equation}
%The detectors are very far from the source. We consider the far zone or wave zone solution of the gravitational wave field.  Which is given by (Ref 1 COmmon reAder)
%\[h_W^{\alpha \beta}=\frac{4GM}{}\]

\subsection{External potentials and their derivatives}
In this sub-section, we derive the post-Newtonian potentials and their derivatives as introduced in Eq. (\ref{acc_eqn}). It is considered that the distance between the two bodies is much larger than the sizes of the two bodies. From here on, external potentials are referred to as just ``potentials".

Before introducing other external potentials and their derivatives, we introduce the following tools which are used to obtain the same% (Because we are differentiating wrt $y_P^j$):
\begin{eqnarray}
    \partial^j r_{SQ} &=&  n_{SQ}^j, \\ 
    \partial_j \mathbf{x}^k_{SQ}&=&  \delta_j^k, \\
    \dot{\mathbf{x}}_{SQ}&=& -\mathbf{v}_Q,\\
    \dot{{r}}_{SQ}&=& -\mathbf{\hat{n}}_{SQ}.\mathbf{v}_Q,  \\
    \partial^jn_{SQ}^k&=&\frac{\delta^{jk}}{r_{SQ}}-\frac{n_{SQ}^jn_{SQ}^k}{r_{SQ}},\\
    \partial^j v_Q^k&=&0.   
\end{eqnarray}

The over-dot represents the time derivative of the quantity. %A substitution has been made $\epsilon^2=l$ for convenience. 
Following are the expressions for the derivatives of the two potentials mentioned in Eq. (\ref{NU}) and Eq. (\ref{NUj}) \cite{Eric_Will}.
\begin{eqnarray}
    \dot{U}_P&=&\frac{G M^*_Q r_{PQ}}{(r_{PQ}^2 +\epsilon^2)^{\delta+1}} (\mathbf{\hat{n}}_{PQ}.\mathbf{v}_Q) ,
    \label{Udot}\\
    \dot{U}^j_P &=&  \frac{G M^*_Q r_{PQ}}{2\delta(r_{PQ}^2+\epsilon^2)^{\delta+1} } \bigg[ 2\delta   (\mathbf{\hat{n}}_{PQ}.\mathbf{v}_Q) v_Q^j  \nonumber\\
    &    &\text{        } +\frac{G M_P^*} {(r_{PQ}^2 + \epsilon^2)^\delta}{n}^j_{PQ}\bigg],
    \label{Uotj} \\
    \partial^j U_P&=&-\frac{G M_Q^* r_{PQ}}{(r_{PQ}^2+\epsilon^2)^{\delta+1}}  n^j_{PQ} ,
    \label{ju} \\ 
    \partial^jU^k_P&=& -\frac{ G M^*_Q r_{PQ}}{ (r_{PQ}^2 + \epsilon^2)^{\delta+1}} {n}^j_{PQ} {v}^k_Q.
    \label{juk}
\end{eqnarray}

Post-Newtonian correction ($\psi_P$) to the Newtonian Potential.
\begin{equation}
    \psi_P=U_P \left(\frac{3}{2}v_Q^2 - U_Q\right)\,\, ,
\end{equation}
where $U_Q$ is the potential at body-Q.
Spatial derivative of $\psi_P$ is
\begin{eqnarray}
\partial^j\psi_P  &=& \partial^j\left[U_P \left(\frac{3}{2}v_Q^2 -U_Q \right) \right] \nonumber\\
    %&\quad =\partial^j\left[k_Q \log(r_{PQ}^2+l)\left(\frac{3}{2}v_Q^2 -k_P \log(r_{QP}^2  +l)\right)\right] \\
    &=& -\frac{3}{2}\frac{ G M^*_Q r_{PQ} }{(r_{PQ}^2 +\epsilon^2)^{\delta+1}} v_Q^2 n_{PQ}^j.
    \label{jpsi}
\end{eqnarray}

Mathematically, potential is the quantity obtained from Poisson's equation using a source term. We need mass density to get the gravitational potential ($U$). Using this Newtonian gravitational potential as source, a higher order of potential called super-potential ($X$) can be obtained. 
In other words super-potential is the potential generated by external potential. Here we will briefly explain the derivation:\newline
The Poisson's equation for super-potential ($X$) is given as
\begin{equation}
\nabla^2 X= 2U.
\end{equation}
The source term contains the potential $U$. The potential due to a mass-$Q$ at a field point ($\mathbf{x'}$) is a function of the distance between the field point and the source mass is
\begin{equation}
    U(\mathbf{x'})= U(\mathbf{x'}-\mathbf{y}_Q).
    \label{U}
\end{equation}
The corresponding super-potential ($X(\mathbf{x}_S)$) at another field point ($\mathbf{x}_S$) due to the potential $U(\mathbf{x'})$ is the solution of the Poisson's equation, $\nabla^2X(\mathbf{x}_S)=2U(\mathbf{x}_S-\mathbf{x'})$. $X(\mathbf{x}_S)$ is spherically symmetric hence it's dependence on vector ($\mathbf{x}_S-\mathbf{x'}$) can be simplified into the dependence over the magnitude of this vector, which we denote as `$r$'. The derivative of this super-potential as required in Eq. (\ref{acc_eqn}) is formalized as
\begin{eqnarray}
    \partial^j\ddot{X} (\mathbf{x_S})\ & =& \partial^j(\partial_{tt} X(r) ) \nonumber\\ 
    & =& \partial^j \left(X''(r) \dot{r}^2 + X'(r) \ddot{r}\right) \nonumber\\ 
    \ & =& X'''(r)\dot{r}^2 \partial^jr  + 2X''_P(r)\dot{r}\partial^j\dot{r}\nonumber \\ 
    && +X''(r)\ddot{r} \partial^jr  + X'(r) \partial^j\ddot{r}.
\end{eqnarray}
Here, the primes denote radial derivative and the over-dots are for time derivatives. The general solution to the Poisson's equation is given as
\begin{equation}
    X(\mathbf{x}_S)=-\frac{1}{2\pi} \int\frac{U(\mathbf{x}')}{|\mathbf{x}_S-\mathbf{x'}|} d^3x' + X_0(\mathbf{x}_S).
    \label{gen sol}
\end{equation}
Here, $X_0$ is the constant of integral, which is the solution to the Laplace equation $\nabla^2 X_0=0$. As prescribed  \cite{Eric_Will} (in the box 7.3) the domain of integration is truncated to spatial near zone. This changes the form of Poisson's equation to $\nabla^2 X= 2U\Theta(R-r)$, where $\Theta(R-r)$ is the Heaviside step function and $R$ is the radius of the boundary of the near zone (value of which is immaterial and this fact will be used later). \\

According to our assumption, the body-Q is heavy enough compared to the body-P such that the effect due to the body-P is negligible, and we consider only the effect due to body-Q as
\begin{equation}
    U(\mathbf{x'})\approx U(\mathbf{x'}-\mathbf{y}_Q) = \frac{GM^*_Q}{2\delta (r_{1Q}^2+\epsilon^2)^{\delta}},
    \label{U_approx}
\end{equation}
here, $r_{1Q}=|\mathbf{x'}-\mathbf{y}_Q|$.\\
Substituting Eq. (\ref{U_approx}) into Eq. (\ref{gen sol}), we obtain
\begin{equation}
    X(\mathbf{x}_S)=-\dfrac{GM^*_Q}{4\pi\delta} \int\frac{\Theta(R-r')}{|\mathbf{x}_S-\mathbf{x'}| (r^2_{1Q}+\epsilon^2)^\delta} d^3x' + X_0(\mathbf{x}_S).
\end{equation}
From here on, $r'=|\mathbf{x'}|$.
If the distance between points, $r_{1Q}$, is very large compared to the scale length $\epsilon$, i.e. $\epsilon << r_{1Q}$, we can approximate the potential as
\begin{equation}
    U(\mathbf{x'-y}_Q) \approx \frac{GM^*_Q}{2\delta (r_{1Q})^{2\delta}}.
\end{equation}
Therefore,
\begin{equation}
    X(\mathbf{x}_S)=-\dfrac{GM^*_Q}{4\pi\delta} \int\frac{\Theta(R-r')}{|\mathbf{x}_S-\mathbf{x'}| (r_{1Q})^{2\delta}} d^3x' + X_0(\mathbf{x}_S).
\end{equation}
For the mathematical simplicity, we take $\mathbf{y}_Q=0$. \newline 
At the end of this section, we generalize our result by introducing $\mathbf{y}_Q$.
We set the Cartesian coordinates such that the Z-axis aligns with the position vector $\mathbf{x_S}$. \newline 
Therefore, $\mathbf{x_S} = r\mathbf{\hat{z}}$ ,  $\mathbf{x_S}.\mathbf{x'} = rr'\cos\theta'$ ,  and $|\mathbf{x}_S-\mathbf{x'}| = \sqrt{r^2 + r'^2 - 2rr'\cos\theta'}$. Hence
\begin{equation}
     X(\mathbf{x}_S)=-\dfrac{GM^*_Q}{4\pi\delta} \int\frac{\Theta(R-r') r'^2\sin\theta'dr'd\theta'd\phi'}{r'^{2\delta}\sqrt{r^2 + r'^2 - 2rr'\cos\theta'}} + X_0.
\end{equation}
The integrand is independent of $\phi'$, thus the $\phi$ integral will give a factor $2\pi$. Now, we integrate $\theta'$ part first. Let us define it as $I(r,r')$.
\begin{eqnarray}
    I(r,r') &=& \int_{0}^{\pi} (r^2 + r'^2 - 2rr'\cos\theta')^{-1/2}\sin\theta' d\theta'\\
            &=& \dfrac{1}{rr'}\{(r+r')-|r-r'| \}
\end{eqnarray}\newline
Now we integrate over $r'$.
\begin{eqnarray}
    X(\mathbf{x}_S) &=&-\dfrac{GM^*_Q}{2\delta} \int_{0}^{R}r'^{2(1-\delta)} I(r,r') dr' + X_0 \nonumber\\
    &=&-\dfrac{GM^*_Q}{2\delta} \bigg(\int_{0}^{r}r'^{2(1-\delta)} \dfrac{2}{r} dr' + \int_{r}^{R}r'^{2(1-\delta)} \dfrac{2}{r'} dr'\bigg) + X_0\nonumber\\
    &=&\dfrac{GM^*_Q}{2\delta(1-\delta)}\bigg(\dfrac{r^{2(1-\delta)}}{3-2\delta} - R^{2(1-\delta)} \bigg) + X_0.
    \label{super1}
\end{eqnarray}
\newline
Now, one should consider $X_0 = \dfrac{GM^*_Q}{2\delta(1-\delta)} R^{2(1-\delta)}$ to get rid of the terms dependent on the boundary of near zone. One can get the following expression of the super-potential
\begin{equation}
    X(\mathbf{x}_S) = \dfrac{GM^*_Q}{2\delta(1-\delta)(3-2\delta)} (r_{SQ})^{2(1-\delta)}\,\, ,
\end{equation}
where we replace $\mathbf{x}$ by $\mathbf{x-y}_Q$ to generalize the result (Eq.~(\ref{super1})) and we drop the assumption $\mathbf{y}_Q=0$.
Placing the body-P at point S, it can be shown that
\begin{widetext}
\begin{equation}
\partial^j \ddot{X}_P = \dfrac{2G M_Q^*}{(3-2\delta)(r_{PQ})^{2\delta+1}} \Bigg[2(\delta+1)(\mathbf{v}_Q . \mathbf{\hat{n}}_{PQ})^2 n^j_{PQ} - v_Q^2 n^j_{PQ} - 2 (\mathbf{v}_Q . \mathbf{\hat{n}}_{PQ}) v_Q^j + \dfrac{2GM^*_P}{(r_{PQ})^{2\delta}} n^j_{PQ} \Bigg].
\label{superpot}
\end{equation}
\end{widetext}
%Using the above expression and Eq. (\ref{acc_eqn}), we determine the acceleration of body-P in the following section.

Acceleration of body-P can be obtained by substituting the expressions for the derivatives of the potentials as derived in Eq. (\ref{Udot}), (\ref{Uotj}), (\ref{ju}), (\ref{juk}), (\ref{jpsi}), and (\ref{superpot}) in Eq. (\ref{acc_eqn}). As we previously mentioned, we assume that the body-Q is much heavier than the body-P. Therefore, for mathematical simplicity, we can consider the body-Q is always stationary which implies $\mathbf{v}_Q \approx 0$ and $\mathbf{a}_Q\approx0$. For simplicity in notations, we use $r_{PQ} = r$, $\mathbf{\hat{n}}_{PQ} = \mathbf{\hat{n}}$, $\mathbf{v}_P=\mathbf{v}$, and $\dfrac{GM_Q^*}{2}=K_Q$. Therefore, the potential and it's differentials reduce to the expressions listed as 

\begin{eqnarray}
     U_P &=& \dfrac{K_Q}{\delta r^{2\delta}} , \label{U_P}\\
     \partial^jU_P &=& \dfrac{-2K_Q}{r^{2\delta+1}} n^j,\\
     U_P^j &=& \dfrac{K_Q v_Q^j}{\delta r^{2\delta}} =0 ,\\
     \dot{U}_P &=& \dfrac{2K_Q (\mathbf{v}_Q.\mathbf{\hat{n}})}{r^{2\delta+1}} = 0 ,\\
     \partial^j\psi_P &=& \dfrac{-3 K_Q v_Q^2}{r^{2\delta+1}} n^j =0 ,\\
    \dot{U}_P^j &=&  \dfrac{2K_PK_Q}{\delta r^{4\delta+1}} n^j + \dfrac{2K_Q(\mathbf{v}_Q.\mathbf{\hat{n}})}{r^{2\delta+1}}v_Q^j  \nonumber \\
     &=& \dfrac{2K_PK_Q}{\delta r^{4\delta+1}} n^j ,\\
     \partial^jU_P^i &=& \dfrac{-2K_Q}{r^{2\delta+1}} n^j v_Q^i= 0. \label{djUi}
\end{eqnarray}
One can substitute the above expressions (Eqs. (\ref{U_P}) to (\ref{djUi})) along with the expression for the derivative of the super-potential as given in Eq. (\ref{superpot}) into Eq. (\ref{acc_eqn}) to obtain the acceleration ($a_P^j$) as follows.
\begin{widetext}
\begin{equation}
    a_P^j = -\dfrac{GM_Q\epsilon^{2\delta-1}}{r^{2\delta+1}} n^j - \dfrac{GM_Q\epsilon^{2\delta-1}}{c^2 r^{2\delta+1}} \bigg[-4 (\mathbf{v}_P.\mathbf{\hat{n}}) v_P^j + \bigg(v_P^2 - \dfrac{2GM_Q\epsilon^{2\delta-1}}{\delta r^{2\delta}} \bigg) n^j \bigg],
    \label{eqn of motion}
\end{equation}
\end{widetext}
where the mass of the body-Q ($M_Q$) is related to the scale mass $M_Q^*$ by $M^*_Q=M_Q \epsilon^{2\delta-1}$.  In the next section, we derive the dynamical variables corresponding to the binary.

\section{THE BINARY SYSTEM AND THE DYNAMICAL QUANTITIES}
\label{three}
The EMRI under consideration comprises of a supermassive compact object (body-Q) and a stellar mass object (body-P) moving in a circular orbit around body-Q.
%For circular orbits at 0-PN order, the radial distance-$r=p$ (constant). 
The position $\mathbf{x}$ of the body-P is written as
\begin{equation}
    \mathbf{x} = r(\cos{\phi} \mathbf{\hat{i}} + \sin{\phi}\mathbf{\hat{j}}),
\end{equation}
where $\phi$ is the azimuthal angle, $r$ is the radius. Since we don't need to work with the dynamics of any body other than body-P, we drop the subscript-``P", i.e. the quantities $r_P$, $\mathbf{x}_P$, $\mathbf{v}_P$ and $\mathbf{a}_P$ will be denoted as $r$, $\mathbf{x}$, $\mathbf{v}$ and $\mathbf{a}$ and the subscript ``PQ" in the quantities that are measured with reference to body-Q are also dropped. i.e. $r_{PQ}$, $\mathbf{x}_{PQ}$ and $\mathbf{\hat{n}}_{PQ}$ are henceforth denoted as $r$, $\mathbf{x}$, $\mathbf{\hat{n}}$. In this study, we have used the following sign convention: $U(r)=+\dfrac{GM_Q\epsilon^{2\delta-1}}{\delta r^{2\delta}}$, and $a(r)=+\dfrac{dU}{dr}$. The differential orbit equation (for the body-P) can be written as \cite{Goldstein,Struck_2015, Struck_2006, Bambhaniya:2019pbr}
\begin{equation}
    \dfrac{d^2u}{d\phi^2}+u=\dfrac{M_P}{L^2}\dfrac{dU(1/u)}{du},
\end{equation}
where $u=1/r$, $L$ is conserved angular momentum of the body-P. Therefore
\begin{equation}
    \dfrac{d^2u}{d\phi^2}+u=\dfrac{GM_QM_P}{L^2} \epsilon^{2\delta-1} u^{2\delta-1}.
\end{equation}
For the circular orbits, the solution of the above differential orbit equation is $u=1/p$. Therefore, we obtain
\begin{equation}
    p^{2(\delta-1)}=\dfrac{GM_QM_P}{L^2} \epsilon^{2\delta-1}.
\end{equation}
\subsection{Dynamical quantities at Newtonian order} \label{NQuantities}
\noindent Henceforth, the dynamical quantities, say F, will appear in the form of $F=F_N + \frac{1}{c^2} F_{PN}$, where $F_N$ denotes the Newtonian order or the 0-PN term which has been derived in this sub-section (\ref{NQuantities}) and $F_{PN}$ denotes the 1PN correction to the quantity F which will be derived in the next sub-section (\ref{PNQuantities}). For example, the Newtonian order and post-Newtonian order expression for velocity of body-P are denoted as $\mathbf{v}_N$ and $\mathbf{v}_{PN}$ respectively while the complete PN-corrected term is denoted as $\mathbf{v}$. 

We derive the Kepler's third law for the power law potential by balancing the centrifugal acceleration with the radial attractive acceleration due to gravity,
\begin{equation}
    \dfrac{v_{N}^2}{r}=\dfrac{GM_Q}{r^{2\delta+1}} \epsilon^{2\delta-1}. \label{Kep3}
\end{equation}
Here, $v_{N}=\dfrac{2\pi r}{T}$ is the 0PN or the Newtonian order expression of the velocity of the body-P and $T$ is the time period of its orbit. Therefore,
\begin{equation}
\dfrac{4\pi^2}{T^2} = \dfrac{GM_Q}{r^{2(\delta+1)}} \epsilon^{2\delta-1}.
\end{equation}
The Kepler's third law in the case of power law potential becomes $T^2 \propto r^{2(\delta+1)}$. We have restricted our attention to the circular orbits, therefore $r=p$ (constant). Thus the angular velocity $\dot{\phi}_N$ becomes
\begin{equation}
    \dot{\phi_{N}}=\dfrac{2\pi}{T}=\sqrt{\dfrac{GM_Q\epsilon^{2\delta-1}}{p^{2(\delta+1)}}}.
\end{equation}
The 0PN velocity $\mathbf{v}_N$ of the body-P using Eq. (\ref{Kep3}) becomes
\begin{equation}
    \mathbf{v}_{N} = \dfrac{d\mathbf{x}}{dt} = \dfrac{d\mathbf{x}}{d\phi} \dot{\phi} =  p\sqrt{\dfrac{GM_Q\epsilon^{2\delta-1}}{p^{2(\delta+1)}}} \boldsymbol{\hat{\nu}}
\label{v0PN}\end{equation}
where $\boldsymbol{\hat{\nu}}=\dfrac{d\mathbf{\hat{n}}}{d\phi}=-\sin{\phi} \mathbf{\hat{i}} + \cos{\phi}\mathbf{\hat{j}}$.\\
The Newtonian order term of the acceleration is
\begin{equation}
    \mathbf{a}_{N} = \dfrac{\partial U}{\partial r}\bigg|_{r=p} = -\dfrac{GM_Q\epsilon^{2\delta-1}}{p^{2\delta+1}} \mathbf{\hat{n}}\,\,.
\end{equation}

\subsection{Dynamical quantities with Post-Newtonian correction} \label{PNQuantities}
\noindent We use the 1PN corrected acceleration (\ref{eqn of motion}) to determine the 1PN correction in other dynamical variables, i.e. velocity, radial coordinate and angular frequency. In the case of circular orbits, $\mathbf{v}_P.\mathbf{\hat{n}}=0$, and the acceleration (\ref{eqn of motion}) reduces to
\begin{equation}
    \mathbf{a} = -\dfrac{GM_Q\epsilon^{2\delta-1}}{p^{2\delta+1}} \mathbf{\hat{n}} + \dfrac{1}{c^2} \dfrac{G^2M_Q^2\epsilon^{2(2\delta-1)}}{\delta p^{4\delta+1}} (2-\delta) \mathbf{\hat{n}},
    \label{simplea}
\end{equation}
which can be written as
\begin{equation}
    \mathbf{a} = \mathbf{a}_N + \dfrac{1}{c^2} \mathbf{a}_{PN},
    \label{defa}
\end{equation}
where the 1PN correction in the acceleration is
\begin{equation}
    \mathbf{a}_{PN} = \dfrac{G^2M_Q^2\epsilon^{2(2\delta-1)}}{\delta p^{4\delta+1}} (2-\delta) \mathbf{\hat{n}}.
\end{equation}
The expression for velocity at the Newtonian order (0PN) can be given as $v_{N}^2 = \dfrac{GM_Q}{p^{2\delta}}\epsilon^{2\delta-1}$ as derived in Eq. (\ref{v0PN}). After substituting this expression, the post-Newtonian parameter ($v_{N}^2/c^2$) becomes evidently apparent. Hence we can simplify Eq. (\ref{simplea}),
\begin{equation}
\mathbf{a} =  -\dfrac{GM_Q\epsilon^{2\delta-1}}{p^{2\delta+1}} \bigg[ 1 - \dfrac{v_N^2}{c^2} \dfrac{(2-\delta)}{\delta } \bigg] \mathbf{\hat{n}}.
\label{acceleration}
\end{equation}
In Kepler-Newton (KN) case, $\delta=1/2$, and the acceleration ($\mathbf{a}_{KN}$) reduces to (\cite{Dionysiou}, \cite{Wagoner})
\begin{equation}
\mathbf{a}_{KN} =  -\dfrac{GM_Q}{p^2} \bigg[1 - 3\dfrac{v_{KN}^2}{c^2}  \bigg] \mathbf{\hat{n}}\,\, ,
\end{equation}
where $v_{KN}^2 = \dfrac{GM_Q}{p}$.

Post-Newtonian corrections are only valid as long as $(\frac{v_N}{c})^2\ll 1$ and the PN correction is much smaller than the Newtonian order part. This can be simplified to give the following condition
\begin{equation}
    \label{PNcondition}
    1\gg \big(\frac{v_N}{c}\big)^2 \big(\frac{2-\delta}{\delta}\big)=\frac{GM_Q \epsilon^{(2\delta-1)}}{p^{2\delta} c^2} \frac{2-\delta}{\delta} \implies \delta > 0.
\end{equation}
We shall consider the cases where $\delta > 0$. 

Now, we proceed to determine the PN corrections in the other dynamical variables.
Let the PN corrected angular momentum per unit mass ($r^2\dot{\phi}$) and PN corrected velocity ($\mathbf{v}_P$) be written as
\begin{eqnarray}
   r^2\dot{\phi} &=& |\mathbf{r}\times\mathbf{v}|= p\sqrt{\dfrac{GM_Q\epsilon^{2\delta-1}}{p^{2\delta}}} (1 + \delta h)\, ,
    \label{angular momentum}\\
    \mathbf{v} &=& \sqrt{\dfrac{GM_Q\epsilon^{2\delta-1}}{p^{2\delta}}} \boldsymbol{\hat{\nu}} + \delta\mathbf{v}\, ,
    \label{vunknown}
\end{eqnarray}
where $p\sqrt{\dfrac{GM_Q\epsilon^{2\delta-1}}{p^{2\delta}}}\delta h$ and $\delta \mathbf{v}$ are the PN correction terms.
As in this paper, we consider circular orbits, for a particular orbit, the angular momentum of the body is constant. Therefore, Eq.~(\ref{angular momentum}) becomes %:
%\begin{equation}
%     p\sqrt{\dfrac{GM_Q\epsilon^{2\delta-1}}{p^{2\delta}}} \dfrac{d\delta h}{dt} = 0 %\implies \delta h = %\text{constant}.
%\end{equation}
%The integration constant required to solve this equation is completely arbitrary, so we can ignore it, $\delta h = 0$. So the 1-PN correction in the angular momentum per unit mass-$r^2\dot{\phi}$ is zero for the circular orbits. So 
\cite{Wagoner},
\begin{equation}
   r^2\dot{\phi} = p\sqrt{\dfrac{GM_Q\epsilon^{2\delta-1}}{p^{2\delta}}}
   \label{L},
\end{equation}
here $r$ and $\dot{\phi}$ are corrected up to 1-PN.\\

Since circular orbits are considered, the radial component of the velocity is zero. Hence the velocity can be written as $v=r\dot{\phi}$. 
Also the relation between acceleration $\mathbf{a}$ and the velocity $\mathbf{v}$ can be given as:
\begin{equation}
    \mathbf{a} = \dfrac{d\mathbf{v}}{dt} = \dfrac{d\mathbf{v}}{d\phi} \dfrac{d\phi}{dt} = \dfrac{d\mathbf{v}}{d\phi} \dot{\phi}.
\end{equation}
Using Eq. (\ref{vunknown}) and Eq. (\ref{acceleration}), the above equation can be re-written as,
\begin{widetext}
\begin{eqnarray}
 -\dfrac{GM_Q\epsilon^{2\delta-1}}{p^{2\delta+1}} \bigg[ 1 - \dfrac{1}{c^2} \dfrac{GM_Q\epsilon^{2\delta-1}}{p^{2\delta}}\frac{(2-\delta)}{\delta} \bigg] \mathbf{\hat{n}} &=& \bigg(\sqrt{\dfrac{GM_Q\epsilon^{2\delta-1}}{p^{2\delta}}} \dfrac{d\boldsymbol{\hat{\nu}}}{d\phi} + \dfrac{d\delta\mathbf{v}}{d\phi} \bigg) \dot{\phi} \nonumber\\
    &=& \bigg(\sqrt{\dfrac{GM_Q\epsilon^{2\delta-1}}{p^{2\delta}}} \dfrac{d\boldsymbol{\hat{\nu}}}{d\phi} + \dfrac{d\delta\mathbf{v}}{d\phi} \bigg) \dfrac{1}{p} \sqrt{\dfrac{GM_Q\epsilon^{2\delta-1}}{p^{2\delta}}} \nonumber\\
    &=& -\dfrac{GM_Q\epsilon^{2\delta-1}}{p^{2\delta+1}} \mathbf{\hat{n}} + \dfrac{1}{p} \sqrt{\dfrac{GM_Q\epsilon^{2\delta-1}}{p^{2\delta}}} \dfrac{d\delta\mathbf{v}}{d\phi}.
\end{eqnarray}
\end{widetext}
Equating the 1-PN terms, we get 
\begin{eqnarray}
    \dfrac{2-\delta}{c^2} \dfrac{G^2M_Q^2\epsilon^{2(2\delta-1)}}{\delta p^{4\delta+1}} \mathbf{\hat{n}} &=& \dfrac{1}{p}\sqrt{\dfrac{GM_Q\epsilon^{2\delta-1}}{p^{2\delta}}} \dfrac{d\delta \mathbf{v}}{d\phi}.     
\end{eqnarray}

    Therefore,
\begin{eqnarray}
     \dfrac{d\delta\mathbf{v}}{d\phi} &=& \dfrac{1}{c^2} \dfrac{(2-\delta)}{\delta} \bigg[\dfrac{GM_Q\epsilon^{2\delta-1}}{p^{2\delta}} \bigg]^{3/2} \mathbf{\hat{n}},\\
    \therefore \delta\mathbf{v} &=& -\dfrac{1}{c^2} \dfrac{(2-\delta)}{\delta} \bigg[\dfrac{GM_Q\epsilon^{2\delta-1}}{p^{2\delta}} \bigg]^{3/2} \boldsymbol{\hat{\nu}}.
\end{eqnarray}

Hence, the velocity $\mathbf{v}_P$ of the body-P corrected up to the first order Post-Newtonian term is
\begin{equation}
   \therefore \mathbf{v}_P = \sqrt{\dfrac{GM_Q\epsilon^{2\delta-1}}{p^{2\delta}}} \bigg[1 - \dfrac{1}{c^2} \dfrac{GM_Q\epsilon^{2\delta-1}}{p^{2\delta}}\frac{(2-\delta)}{\delta} \bigg] \boldsymbol{\hat{\nu}}.
    \label{Velocity upto 1PN}
\end{equation}
Substituting Eq. (\ref{Velocity upto 1PN}) in the following relation
\begin{equation}
    \mathbf{v}_P = \dfrac{d\mathbf{x}}{dt} = \dfrac{d\mathbf{x}}{d\phi} \dfrac{d\phi}{dt} = r\dot{\phi} \boldsymbol{\hat{\nu}} = \dfrac{r^2\dot{\phi}}{r}\boldsymbol{\hat{\nu}},
\end{equation}
and using Eq.~(\ref{L}), we obtain
\begin{widetext}
\begin{equation}
    \sqrt{\dfrac{GM_Q\epsilon^{2\delta-1}}{p^{2\delta}}} \bigg[1 - \dfrac{1}{c^2} \dfrac{GM_Q\epsilon^{2\delta-1}}{p^{2\delta}}\frac{(2-\delta)}{\delta} \bigg] \boldsymbol{\hat{\nu}} \\ = \dfrac{r^2\dot{\phi}}{r} \boldsymbol{\hat{\nu}}
    = \dfrac{p}{r} \sqrt{\dfrac{GM_Q\epsilon^{2\delta-1}}{p^{2\delta}}} \boldsymbol{\hat{\nu}},
\end{equation}
\end{widetext}
which implies
\begin{equation}
   \dfrac{p}{r} = \bigg[1 - \dfrac{1}{c^2} \dfrac{GM_Q\epsilon^{2\delta-1}}{p^{2\delta}}\frac{(2-\delta)}{\delta} \bigg].
\end{equation}
Therefore, the radial coordinate ($r$) corrected up to the first order Post-Newtonian term is
\begin{equation}
    r=p\bigg[1+\dfrac{1}{c^2}\dfrac{GM_Q\epsilon^{2\delta-1}}{p^{2\delta}}\frac{(2-\delta)}{\delta} \bigg] + O(c^{-4}).
    \label{r upto 1PN}
\end{equation}
Next, we derive the expression for the angular velocity $\dot{\phi}$. Substituting Eq. (\ref{r upto 1PN}) into Eq. (\ref{L}) and simplifying for $\dot{\phi}$, we obtain
\begin{equation}
    \dot{\phi}=\dfrac{1}{p}\sqrt{\dfrac{GM_Q\epsilon^{2\delta-1}}{p^{2\delta}}} \bigg[1-\dfrac{1}{c^2}\dfrac{2GM_Q\epsilon^{2\delta-1}}{\delta p^{2\delta}} (2-\delta) \bigg].
    \label{dot phi upto 1PN}
\end{equation}
In the next section, we discuss the average energy radiation rate of EMRI using the expressions of PN corrected quantities derived here.

\section{Average Energy radiation rate from an EMRI}
\label{four}
The average energy radiation rate $\langle\frac{dE}{dt}\rangle$ from a binary is given by the following expression \cite{Wagoner, Dionysiou, Eric_Will}.
\begin{equation}
    \bigg \langle \frac{dE}{dt} \bigg \rangle  = -\frac{G}{45c^5} \frac{1}{T} \int^T_0 (\dddot{D}^{\alpha \beta})^2 dt,
    \label{D1}
\end{equation}
where the mass quadrupole tensor $D^{\alpha\beta}$ is defined as \cite{Dionysiou}
\begin{equation}
    D^{\alpha\beta} = M_P (3x^\alpha x^\beta - |\mathbf{x}|^2 \delta^{\alpha\beta}).
\end{equation}
Therefore,
\begin{widetext}
\begin{eqnarray}
    \dddot{D}^{\alpha\beta} &=& \dfrac{d^3}{dt^3} \big[M_P(3x^\alpha x^\beta - |\mathbf{x}|^2 \delta^{\alpha\beta}) \big]\nonumber \\
    &=& M_P \dfrac{d^2}{dt^2} \big[3\dot{x}^\alpha x^\beta + 3x^\alpha \dot{x}^\beta - 2\mathbf{x}.\dot{\mathbf{x}} \delta^{\alpha\beta} \big]\nonumber \\
    &=& M_P \dfrac{d}{dt} \big[3\ddot{x}^\alpha x^\beta + 6\dot{x}^\alpha \dot{x}^\beta + 3x^\alpha\ddot{x}^\beta - 2\dot{x}^2\delta^{\alpha\beta} - 2\mathbf{x}.\ddot{\mathbf{x}}\delta^{\alpha\beta} \big]\nonumber \\
    &=& M_P\big[3\dddot{x}^\alpha x^\beta + 9\ddot{x}^\alpha \dot{x}^\beta + 9\dot{x}^\alpha \ddot{x}^\beta + 3 x^\alpha \dddot{x}^\beta - 6 \dot{\mathbf{x}}.\ddot{\mathbf{x}}\delta^{\alpha\beta} - 2 \mathbf{x}.\dddot{\mathbf{x}}\delta^{\alpha\beta} \big]
\end{eqnarray}
\end{widetext}
Hence, $(\dddot{D}^{\alpha\beta})^2$ can be written as
\begin{widetext}
\begin{eqnarray}
    (\dddot{D}^{\alpha\beta})^2 &=& (\dddot{D}^{\alpha\beta}) (\dddot{D}_{\alpha\beta}) \nonumber\\
    &=& 6M_P^2 \big[3\dddot{x}^2 x^2 + 18(\dddot{\mathbf{x}}.\ddot{\mathbf{x}})(\dot{\mathbf{x}}.\mathbf{x}) + 18(\dddot{\mathbf{x}}.\dot{\mathbf{x}})(\ddot{\mathbf{x}}.\mathbf{x}) + (\mathbf{x}.\dddot{\mathbf{x}})^2 + 27\dot{\mathbf{x}}^2\ddot{\mathbf{x}}^2 + 9(\dot{\mathbf{x}}.\ddot{\mathbf{x}})^2 -12(\dddot{\mathbf{x}}.\mathbf{x})(\ddot{\mathbf{x}}.\dot{\mathbf{x}}) \big],
\label{D^2}
\end{eqnarray}
\end{widetext}
where the acceleration terms ($\ddot{\mathbf{x}}$) can be written as the gradient of the potential or the magnitude of the force corresponding to this acceleration is given as $\mathbf{F_a}=M_P |\mathbf{a}|=\dfrac{\partial V}{\partial r}$, where $V$ is the potential energy. Therefore, the acceleration of the body-P becomes
\begin{equation}
  \mathbf{a} =\frac{1}{M_P} \dfrac{\partial V}{\partial r} \mathbf{\hat{n}} + \dfrac{1}{c^2} \mathbf{a}_{PN}.
\end{equation}
Substituting the above expression of $\mathbf{a}$ into Eq. (\ref{D^2})  and after some simplification, we obtain the following expression for $(\dddot{D}^{\alpha\beta})^2$ (\cite{Dehen, Dionysiou}):
\begin{widetext}
\begin{eqnarray}
    (\dddot{D}^{\alpha\beta})^2 = 24\Bigg[\bigg(r\dot{r}\dfrac{\partial^2 V}{\partial r^2} + 3\dot{r}\dfrac{\partial V}{\partial r} \bigg)^2 &+& 12r^2\dot{\phi}^2\bigg(\dfrac{\partial V}{\partial r} \bigg)^2 \Bigg] + \dfrac{24M_P}{c^2} \Bigg[2r\bigg(\dot{r}\dfrac{\partial^2 V}{\partial r^2} \mathbf{x} + 3\dfrac{\partial V}{\partial r}\dot{\mathbf{x}} \bigg).\dot{\mathbf{a}}_{PN} \nonumber \\ &+& \bigg\{9\dfrac{\partial V}{\partial r}\big(\dot{r}^2+2r^2\dot{\phi}^2 \big)\mathbf{\hat{n}} + 3\dot{r}\dfrac{\partial^2 V}{\partial r^2}\big(3\dot{r}\mathbf{x} - r\dot{\mathbf{x}} \big) + 9\dot{r}\dfrac{\partial V}{\partial r}\dot{\mathbf{x}} \bigg\}.\mathbf{a}_{PN} \Bigg],
    \label{D^2 complete}
\end{eqnarray}
\end{widetext}
For circular orbits $\dot{r}=0$, therefore Eq. (\ref{D^2 complete}) reduces to the following expression.
\begin{eqnarray}
    (\dddot{D}^{\alpha \beta})^2 &=&24\Bigg[12\bigg(\dfrac{\partial V}{\partial r}\bigg)^2r^2\dot{\phi}^2 \nonumber \\ &+& \dfrac{M_P}{c^2}\bigg\{18^2\dot{\phi}^2\dfrac{\partial V}{\partial r}\mathbf{\hat{n}}.\mathbf{a}_{PN} + 6r\dfrac{\partial V}{\partial r}\mathbf{v}_P.\dot{\mathbf{a}}_{PN} \bigg\}    \Bigg].\nonumber\\
    \label{D^2 for circular orbits}
\end{eqnarray}
\newline
Here, the time derivative of the PN correction term $\dot{\mathbf{a}}_{PN}$ of the acceleration is
\begin{equation}
    \dot{\mathbf{a}}_{PN} = \dfrac{d\mathbf{a}_{PN}}{dt} = \dfrac{d\mathbf{a}_{PN}}{d\phi} \dfrac{d\phi}{dt} = \dfrac{d\mathbf{a}_{PN}}{d\phi}\dot{\phi}.
\end{equation}
Therefore,
\begin{equation}
    \dot{\mathbf{a}}_{PN} = \bigg[\dfrac{GM_Q\epsilon^{2\delta-1}}{p^{2\delta}} \bigg]^{5/2} \dfrac{(2-\delta)}{\delta p^2} \boldsymbol{\hat{\nu}} + O(c^{-2}),
\end{equation}
where the higher order terms are ignored as they will contribute to the 2PN or higher orders, which is out of the scope of this current study. We can also write,
\begin{eqnarray}
     V(r) &=& \dfrac{GM_QM_P}{2\delta r^{2\delta}}\epsilon^{2\delta-1}, \\
     \mathbf{v}_P.\dot{\mathbf{a}}_{PN} &=& \dfrac{G^3M_Q^3\epsilon^{3(2\delta-1)}}{\delta p^{2(3\delta+1)}}(2-\delta) + O(c^{-2}),\\ 
    \mathbf{\hat{n}}.\mathbf{a}_{PN} &=& \dfrac{G^2M_Q^2\epsilon^{2(2\delta-1)}}{\delta p^{4\delta+1}}(2-\delta) + O(c^{-2}).
\end{eqnarray}
\newline
Now substituting Eq. (\ref{r upto 1PN}), (\ref{dot phi upto 1PN}), and the above expressions into Eq. (\ref{D^2 for circular orbits}), we obtain
\begin{widetext}
\begin{equation}
    (\dddot{D}^{\alpha \beta})^2 = \dfrac{288G^3M_Q^3M_P^2\epsilon^{3(2\delta-1)}}{p^{2(3\delta+1)}} \bigg[1-\dfrac{1}{c^2} \dfrac{2GM_Q\epsilon^{2\delta-1}}{\delta p^{2\delta}}(2-\delta)(3+2\delta) \bigg].
\end{equation}
\end{widetext}

The average rate of the energy radiation mentioned in Eq.~(\ref{D1}) can be written as,

\begin{eqnarray}
       \bigg \langle \frac{dE}{dt} \bigg \rangle 
        = -\frac{G}{45c^5} \frac{1}{2\pi} \int^{2\pi}_0 (\dddot{D}^{\alpha \beta})^2 d\phi\,\, ,
        \label{D2}
\end{eqnarray}
where $\dot{\phi}=2\pi/T$.
\begin{figure*}
\centering
\subfigure[For $\delta=0.27~( i.e.~\delta< \delta_0)$]
{\includegraphics[width=82mm]{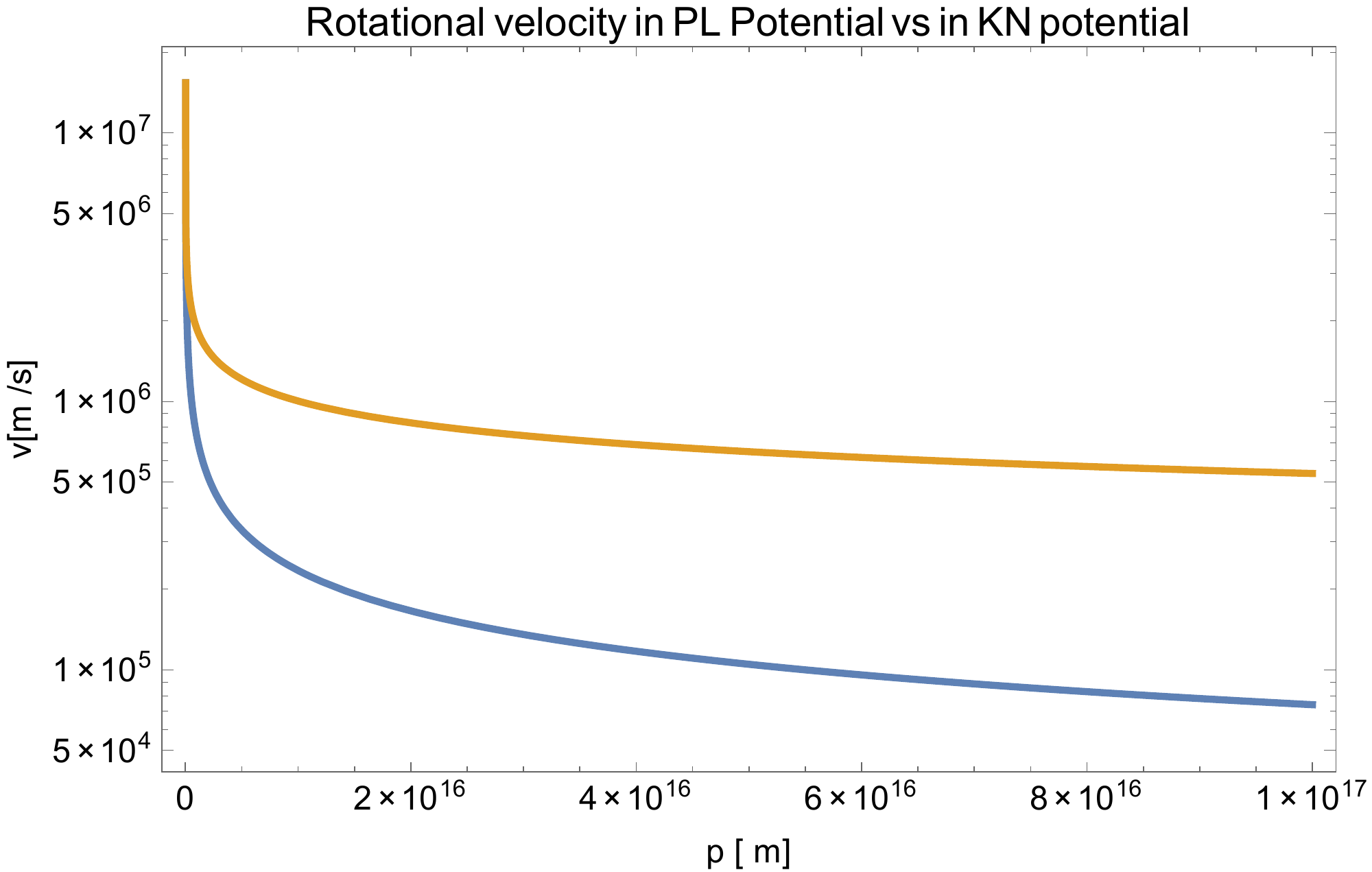}\label{Vdelta0.27_(1)}}
\subfigure[For $\delta = 0.01 \text{ i.e. } \approx 0$]
{\includegraphics[width=82mm]{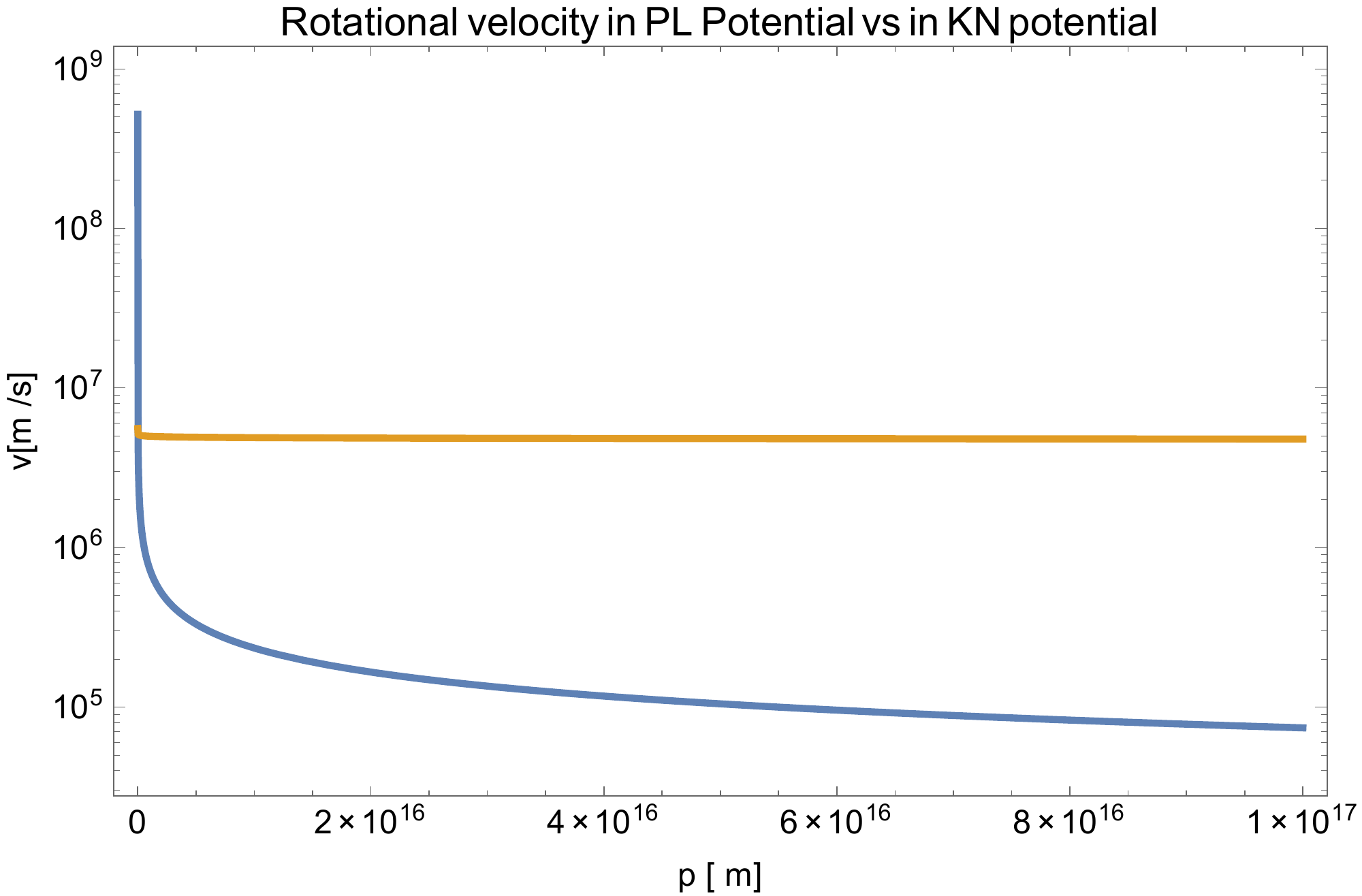}\label{Vdelta0.01_(1)}}
\hspace{0.2cm}
\subfigure[For $\delta=0.535~(i.e.~ \delta > 0.5)$]
{\includegraphics[width=85mm]{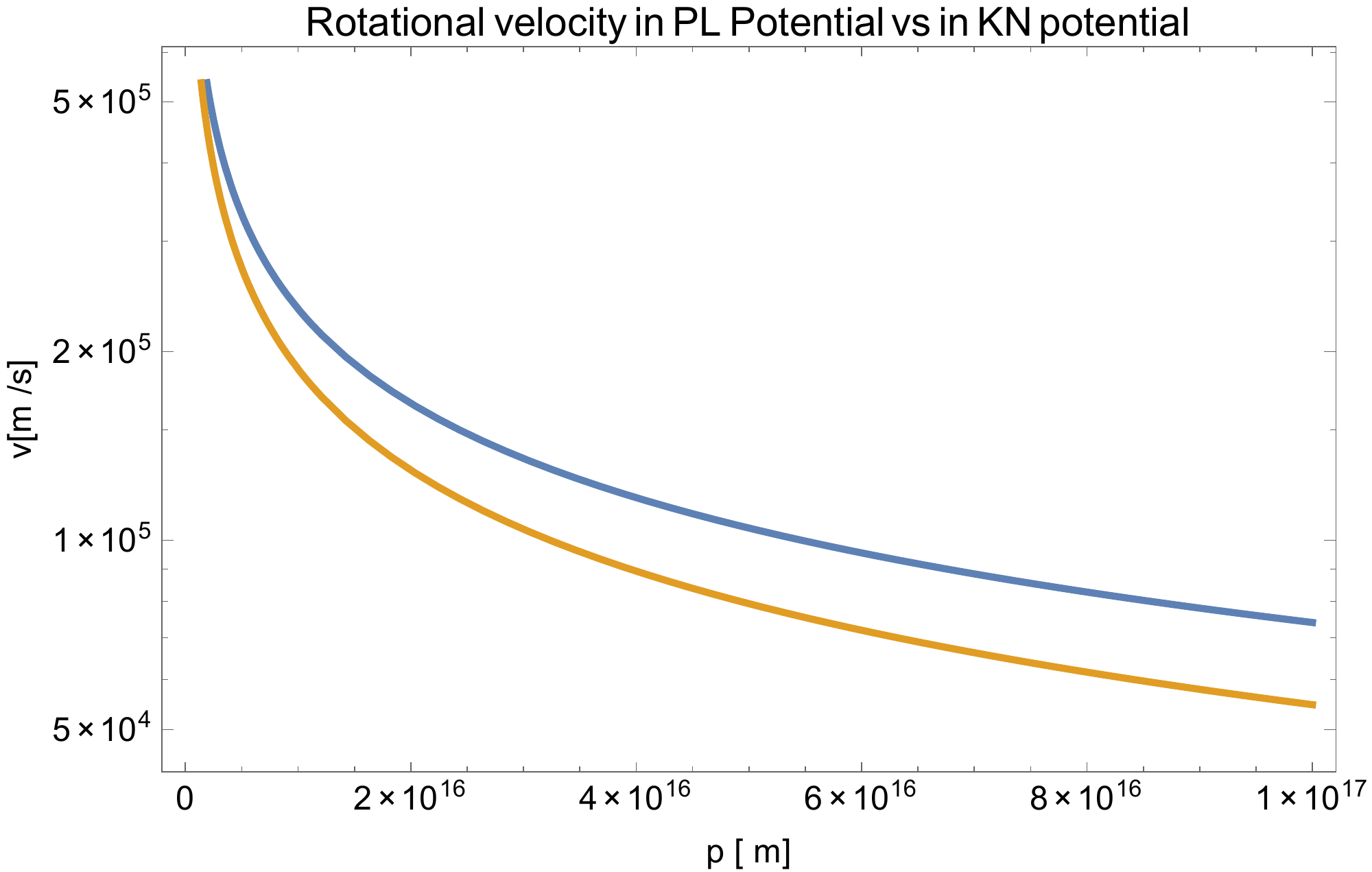}\label{Vdelta0.535_(1)}}
\hspace{0.2cm}
\caption{Figure shows velocity of the smaller object in the  extreme mass ratio binary in power law (PL) potential (Orange curve) and in Kepler-Newton (KN) potential (Blue curve) with respect to radial distance between the two bodies.}
\label{VFigs}
\end{figure*}
In the case of circular orbits, the integrand is independent of the variable $\phi$. Therefore, the above expression of the average energy radiation rate reduces to

\begin{equation}
      \bigg \langle \frac{dE}{dt} \bigg \rangle  = -\dfrac{G}{45c^5} (\dddot{D}^{\alpha\beta})^2.
  \end{equation}\newline
Thus
\begin{widetext}
\begin{equation}
    \bigg \langle \frac{dE}{dt} \bigg \rangle = -\dfrac{32G^4M_Q^3M_P^2\epsilon^{3(2\delta-1)}}{5c^5p^{2(3\delta+1)}} \bigg[1-\dfrac{1}{c^2}\dfrac{2GM_Q\epsilon^{2\delta-1}}{\delta p^{2\delta}}(2-\delta)(3+2\delta) \bigg].
    \label{energy radiation result}
\end{equation}
\end{widetext}
The condition for using PN approximations to evaluate average energy radiation rate for power law potential is
\begin{equation}
    1\gg \frac{G M_Q}{c^2 p^{2\delta}} \frac{(2-\delta)(3+2\delta)}{\delta}.
\end{equation}
For a given EMRI and in a range of radial distance, this condition can be narrowed down to a cut-off  $\delta_0$ below which PN approximations cannot be used to work out the average energy radiation rate. This cut-off depends on the properties of the binary under consideration.
\begin{figure*}[t]
\centering
\subfigure[$\delta=0.27~( i.e.~\delta< \delta_0)$]
{\includegraphics[width=82mm]{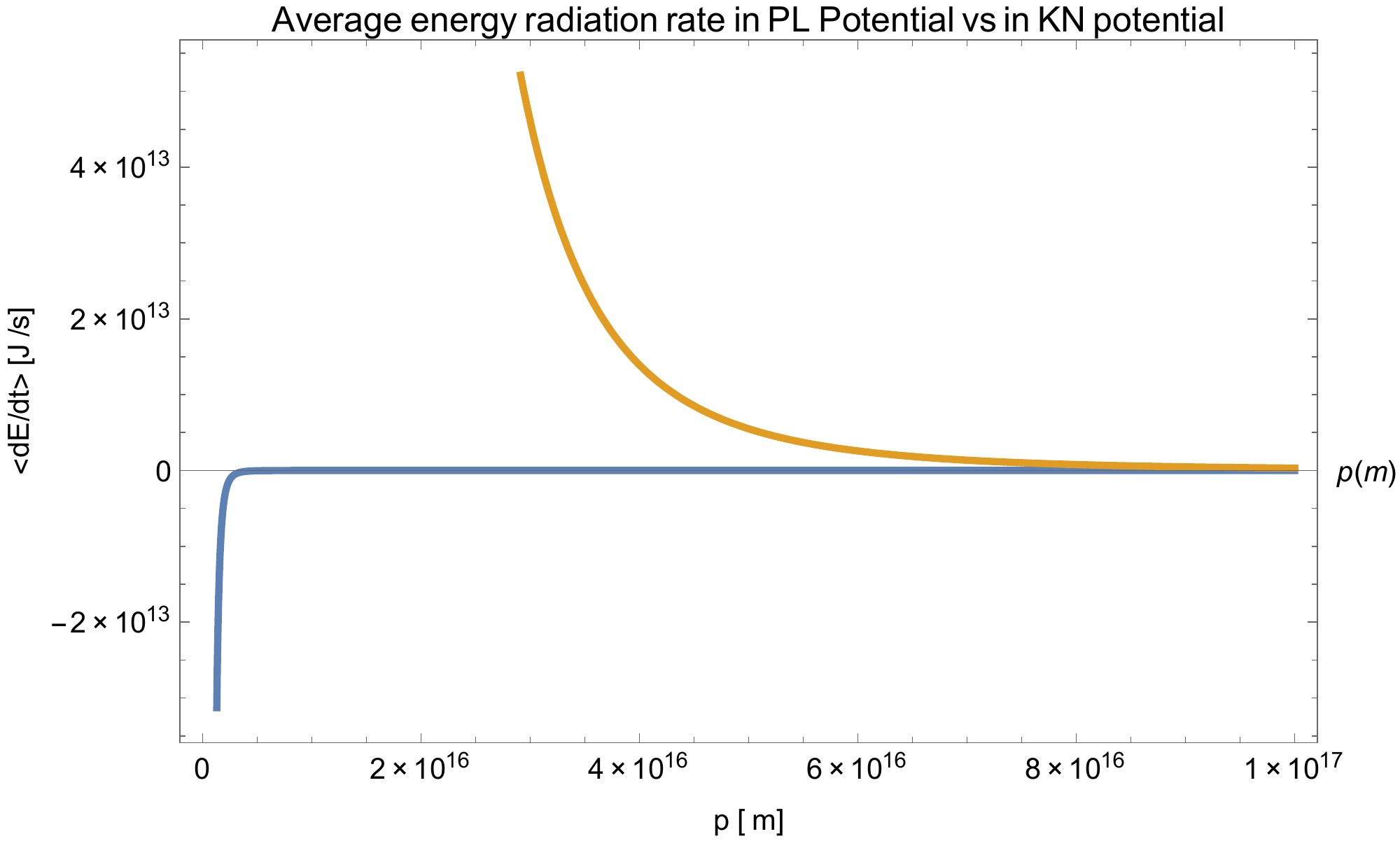}\label{delta0.27_(1)}}
\hspace{0.2cm}
\subfigure[$\delta=0.395 ~(i.e.~ \delta_0<\delta < 0.5)$]
{\includegraphics[width=82mm]{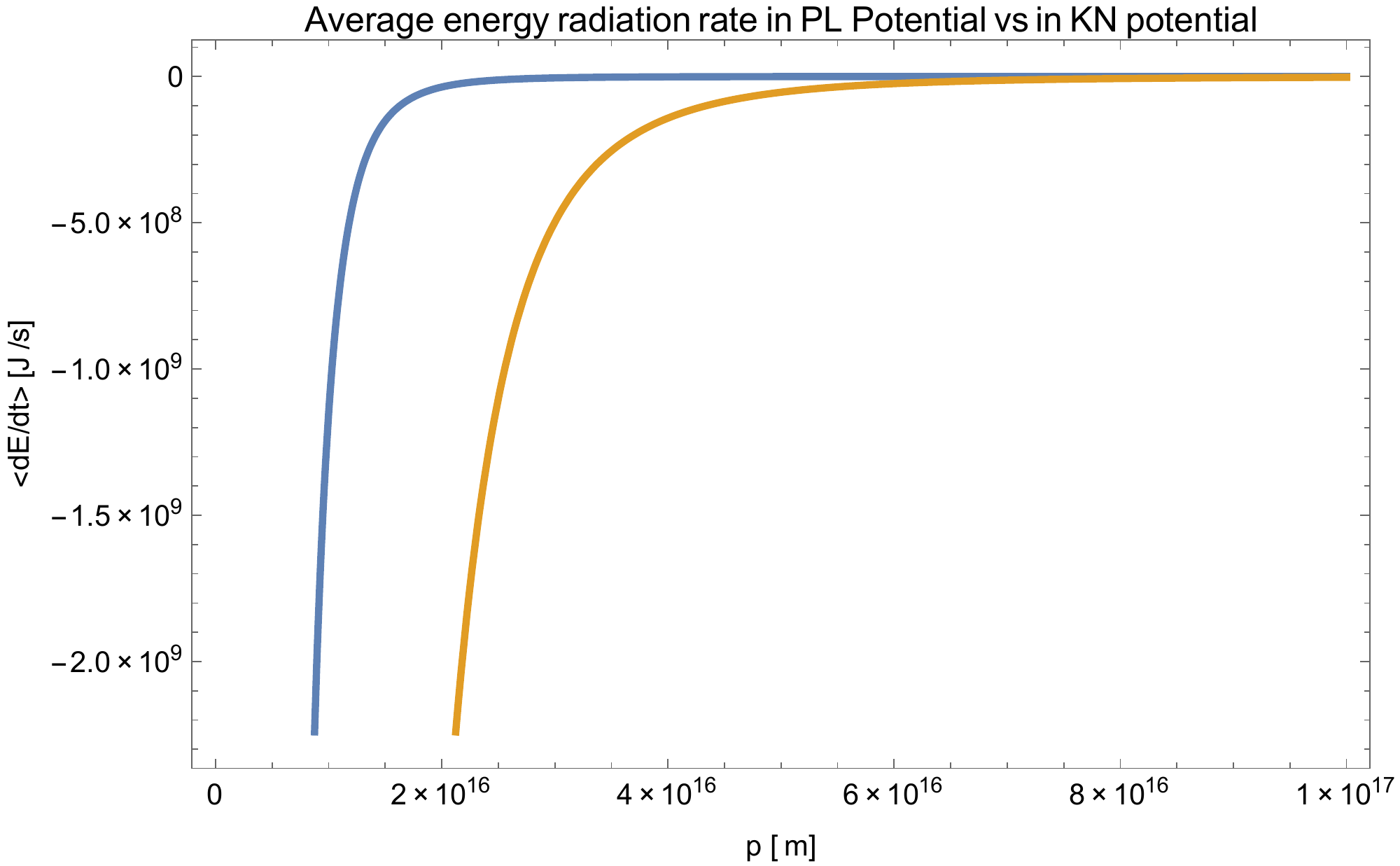}\label{delta0.395}}\\
\subfigure[$\delta=0.535~(i.e.~ \delta > 0.5)$]
{\includegraphics[width=85mm]{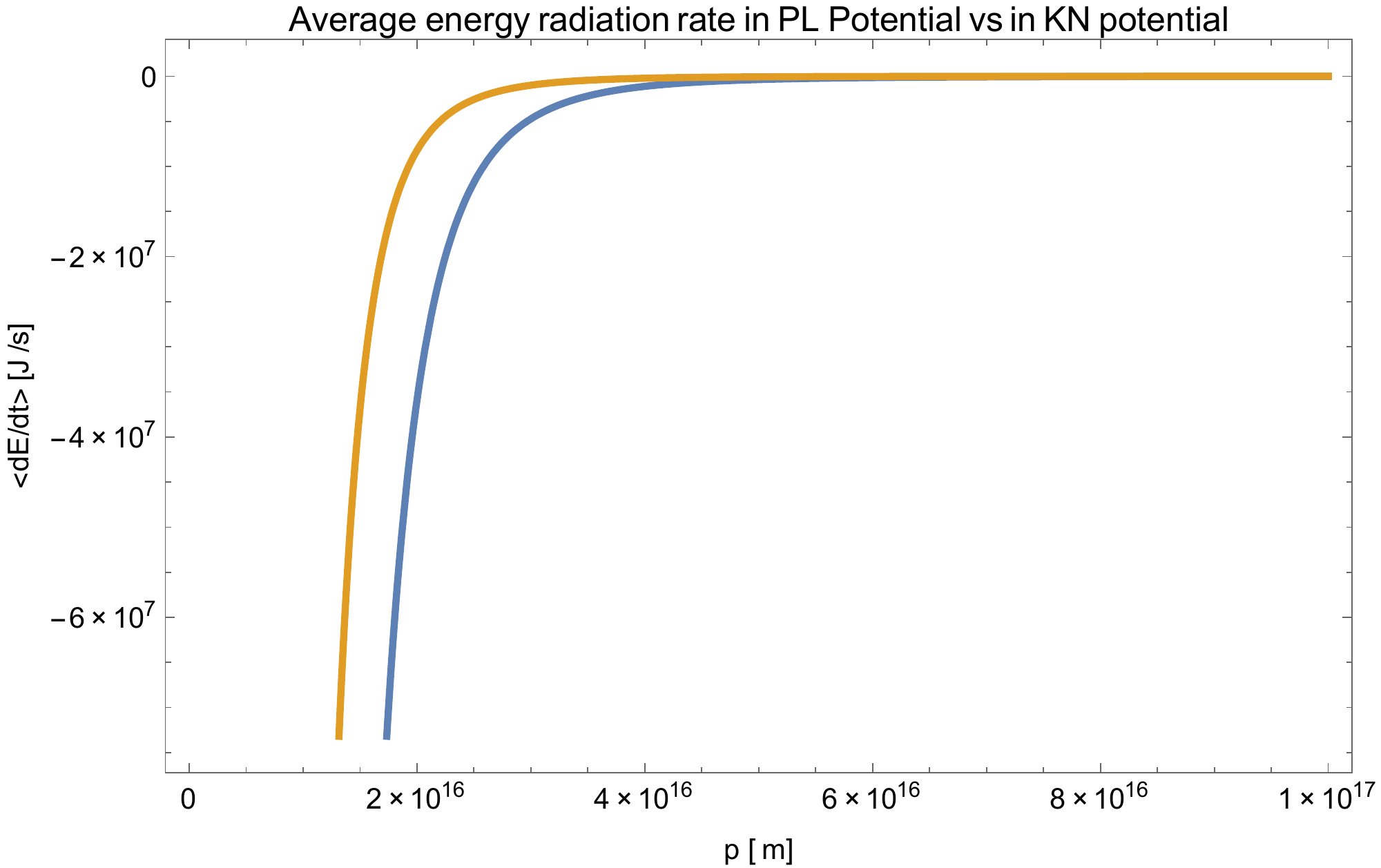}\label{delta0.535}}
\hspace{0.2cm}
\caption{Figure shows variation of 1-PN corrected rate of energy radiation from an extreme mass ratio binary in power law (PL) potential (Orange curve) and in Kepler-Newton (KN) potential (Blue curve)  with respect to radial distance  ($p$)  between the two bodies.}
\label{Efigs}
\end{figure*} 
\begin{figure*}[t]
\centering
\subfigure[$\delta=0.395 ~(i.e.~ \delta_0<\delta < 0.5)$,  $\Delta=0.105$]
{\includegraphics[width=82mm]{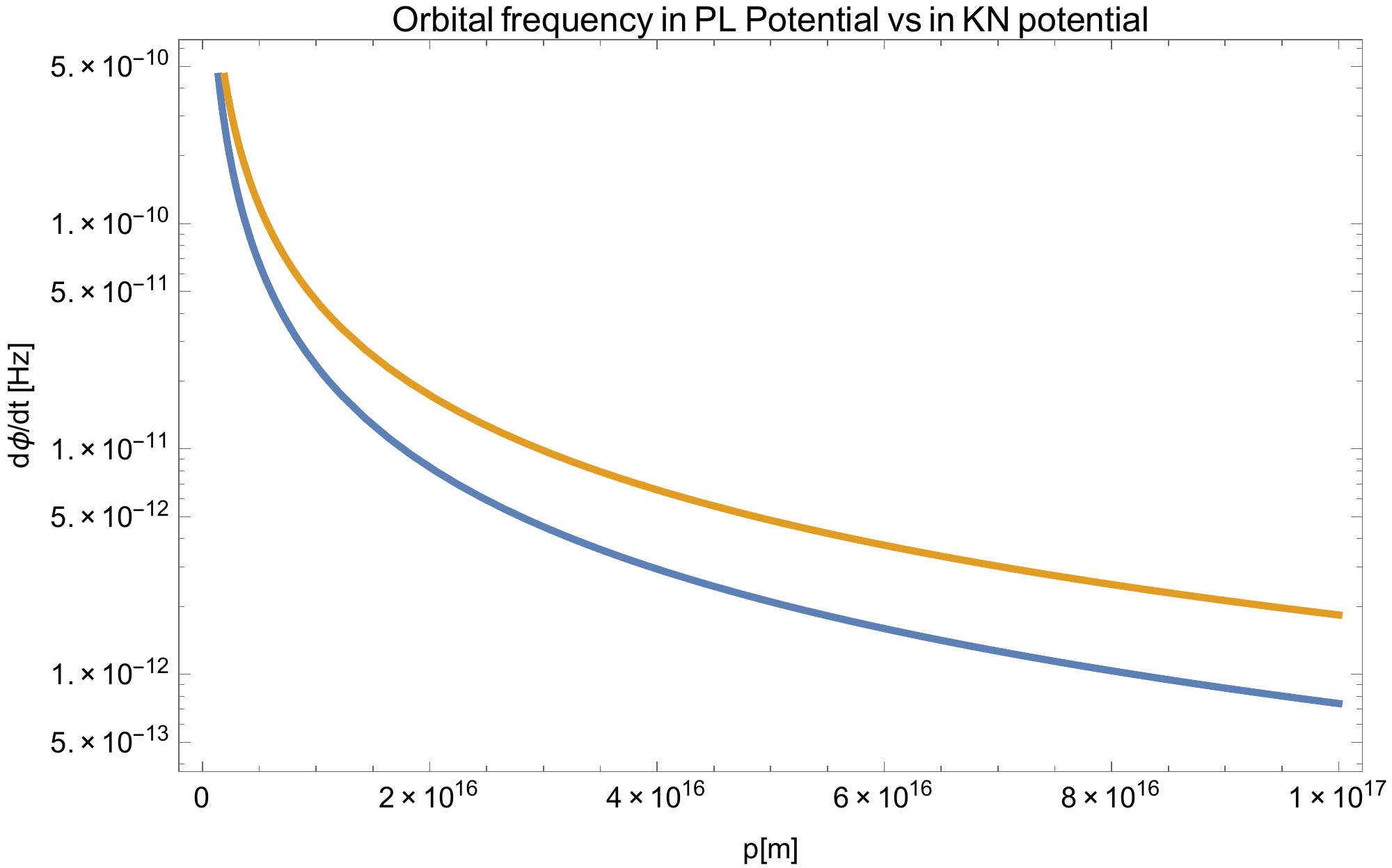}
\label{phidelta0.395}}
\subfigure[$\delta=0.535~(i.e.~ \delta > 0.5),~\Delta=-0.035$]
{\includegraphics[width=82mm]{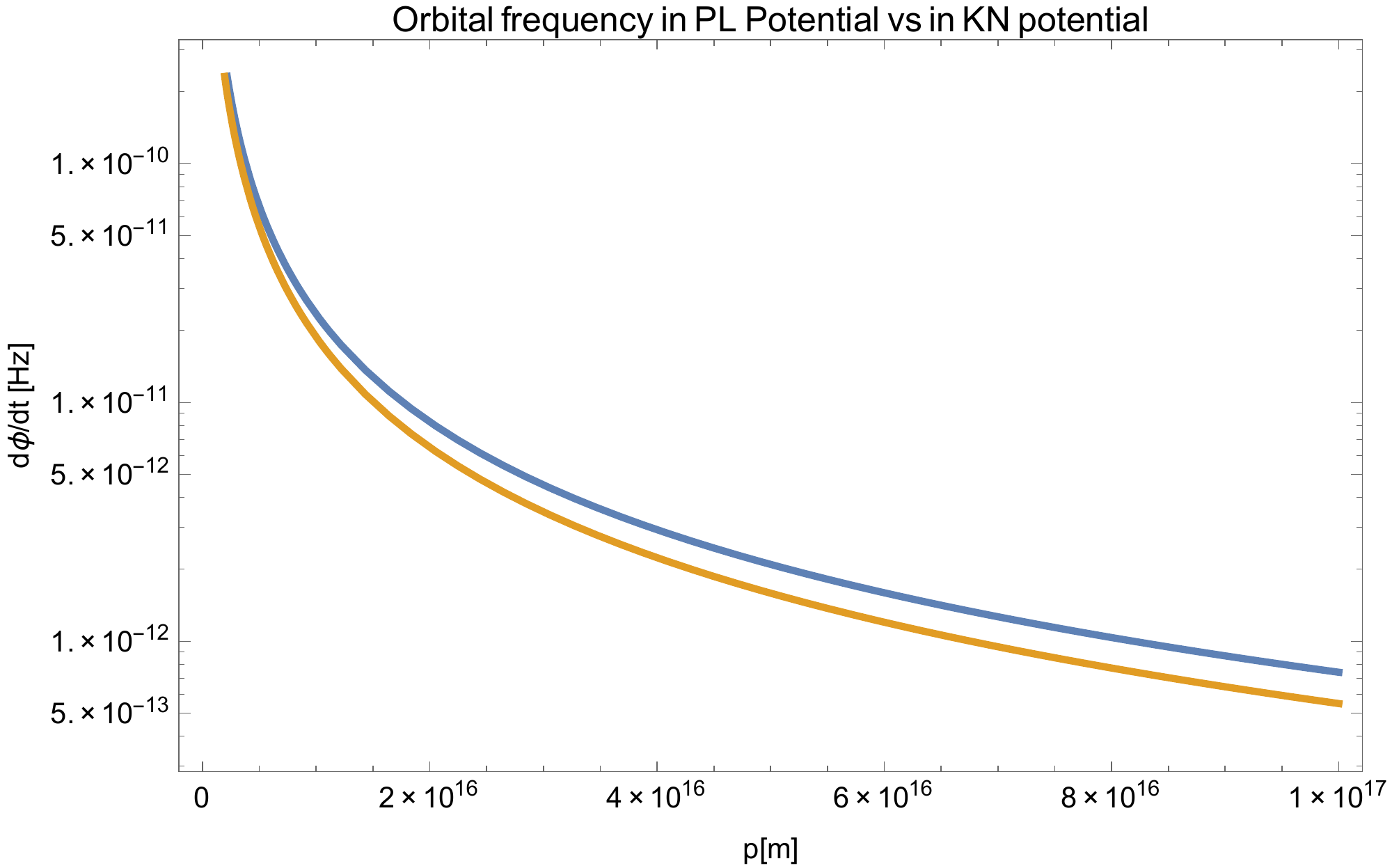}\label{phidelta0.535}}
\hspace{0.2cm}
\caption{Figure shows the comparison of the orbital frequency of the extreme mass ratio binary in power law (PL) potential (Orange curve) and in Kepler-Newton (KN) potential (Blue curve) with respect to radial distance ($p$) between the two bodies.}
\label{Phifigs}
\end{figure*} 

Eq. (\ref{energy radiation result}) shows the energy radiation from the binaries following power law potential with the assumptions that (i) the scale length ($\epsilon$) is very small compared to the distance between two bodies i.e. $r_{PQ}$, (ii) the body-Q is much heavier than the body-P, therefore the motion of the body-Q is negligible, and it can be treated as a stationary body and (iii) the body-P is moving in the circular orbit around Q.

\section{ Comparison of power Law potential with the Kepler-Newton Potential}
\label{five}
Consider an example of an EMRI system of a supermassive black hole, and a star as massive as the sun.\\
Let the properties of the supermassive compact object be: mass, $M_Q$ = $4.1 \times 10^6 M_{\odot} = 8.2 \times 10^{36} kg$, the scale radius, $\epsilon$= $5.8 \times 10^{-4} pc$ and the properties of the star or the stellar mass object be: mass, $M_P$ = $1 M_{\odot}$ = $2 \times 10^{30} kg$.  
%The maximum speed of information transfer, $c$ = $3 \times 10^8$ m/s.
%The Universal gravitational Constant, $G = 6.67 \times 10^{-11} m^3 kg^{-1} s^{-2}$.\\

In this section, we compare the velocity curves, the average energy radiation rate and the orbital frequency of the binary in the PL potential and in the KN Potential. The PN corrections are much smaller than the Newtonian values of the evaluated quantities. Hence, in the magnitude scales considered below, the graphs of the Newtonian and the PN corrected quantities will apparently overlap. 
 
\subsection{Velocity profile}
The comparison of the velocity curves for power law potential (Eq.~(\ref{Velocity upto 1PN})) and Kepler-Newton potential is shown in this subsection. 
For $0<\delta < 0.5$, the velocity falls more slowly in the case of power law (PL) potential as compared to that in KN potential case which is usually observed in astrophysical scenarios (Fig. \ref{Vdelta0.27_(1)}). In fact, for $\delta = 0.01 \text{ i.e. } \approx 0$, the velocity curve is nearly flat (Fig. \ref{Vdelta0.01_(1)}). For the case of $0.5<\delta$, the velocity for PL potential drops faster with increasing radial distance (Fig. \ref{Vdelta0.535_(1)}). Therefore, $0.5<\delta$ is astrophysically improbable.

\subsection{Average energy radiation rate}
Next, we compare the average energy radiation rate from an EMRI in PL potential and that in KN Potential.

Average energy radiation rate from a system in KN potential can be determined by substituting $\delta=1/2$ in our final result (Eq. (\ref{energy radiation result})) 
\begin{equation}
\bigg\langle\frac{dE}{dt}\bigg\rangle_{KNP} = -\frac{32G^4 M_Q^3 M_P^2}{5 c^5 p^5} \bigg(1- \dfrac{1}{c^2} \dfrac{24GM_Q}{p}\bigg).
\label{energy radiation KN}
\end{equation}
The comparisons of the two energy radiation rates, Eq. (\ref{energy radiation KN}), and Eq.~(\ref{energy radiation result}) with $\delta\neq 0.5$, are shown in the Fig.~(\ref{Efigs}) for different $\delta$-s. 

In the EMRI considered above, we focus on the region where the radial distance $p\approx10^{16}$ meter and the corresponding $\delta_0\approx0.38$.
For $0<\delta < 0.38$, the 1-PN term becomes comparable and much higher than the 0-PN (or the Newtonian) part. Therefore, the 1-PN part blows up and it looks as if the EMRI is absorbing energy instead of radiating (Fig. \ref{delta0.27_(1)}). Hence, as far as the energy radiation rate is concerned, we have to limit our study to $\delta > \delta_0 :=0.38$. For $0.38< \delta < 0.5$, the  energy radiation rate is greater than that from KN potential (Fig. \ref{delta0.395}). For $\delta > 0.5$, the energy radiation rate is smaller than that from KN potential (Fig. \ref{delta0.535}). However, as we mentioned before, $\delta>0.5$ is astrophysically improbable.

%\begin{figure}
%\includegraphics[width=1.0\linewidth]{ELPvsKNPdelta345.PDF} 
%\caption{For $0.33<\delta < 0.38$: Variation of 1-PN Correction to rate of energy radiation from an extreme mass ratio binary in power law (PL) potential (Orange curve) and in Kepler-Newton (KN) potential (Blue curve) with respect to radial distance $p$  between the two bodies. In the average energy radiation term the condition: "1PN part $\ll $ 0PN part" is violated. And hence the EMRI seems to absorb energy instead of radiate it.}
%\caption{Caption for this figure with two images}
%\label{delta0.345}
%\end{figure}

\subsection{Orbital frequency}

Presence of matter distribution around a binary affects the dynamics of the binary. The effect of dynamical friction as studied in \cite{bekensteinDynFric, gomezDynfric,CaputoDynFric} is ignored here. 
Let's consider a power law (where the power is $\delta$) such that it has a small deviation ($\Delta$) in the power from the Kepler-Newton potential ($\delta_{KN}=1/2$) which could be because of the presence of matter distribution around the EMRI. Accordingly,  from the Eq.~(\ref{dot phi upto 1PN}), the orbital frequency $\dot{\phi}$ can be written as 
\begin{equation}
     \dot{\phi}=\dfrac{1}{p}\sqrt{\dfrac{GM_Q\epsilon^{1-2\Delta}}{p^{(1-2\Delta)}}} \bigg[1-\dfrac{1}{c^2}\dfrac{2GM_Q\epsilon^{-2\Delta}}{(1/2-\Delta) p^{(1-2\Delta)}} (3/2+\Delta) \bigg].
\end{equation}

When $\Delta>0$, which is also the case where we expect non-vacuum environment, the orbital frequency is increased (Fig. \ref{phidelta0.395}). While in the case where $\Delta<0$ we get orbital frequency lower than that of an EMRI in KN potential but this case astrophysically improbable. 

From the observation data of the orbital frequency of EMRI, the value of $\Delta$ can be obtained which gives the precise power law potential which can be used to get the matter density profile around the central supermassive compact object. 

\section{Conclusion}
\label{six}
In this paper, we consider an extreme mass ratio inspiral (EMRI) in a system where the supermassive object lies in a non vacuum region such that we can use power law potential. The main motivation to consider power law potential is to get nearly flat profile of orbital velocity which is observed in astrophysical scenarios (e.g. galactic rotation curve). We derive the 1-PN corrected dynamical variables of the orbiting stellar mass object. We consider circular orbits for the ease of calculations. Using the Newtonian gravitational potential ($U$) and the PN potentials ($U^j,$ $\Psi$ and $X$), we derive  the acceleration of a body in the PL potential. Using this, we derive other PN corrected dynamical variables. Subsequently, we obtain the mass quadrupole tensor which is ultimately used to obtain the average energy radiation rate.\newline

Next, we discuss an example of an EMRI. We calculate the average energy radiation rate from this EMRI in power law potential at different $\delta$-s and compared them with that from the same EMRI at $\delta=0.5$ i.e. the KN potential case. When $\delta$ is less than a particular cut-off $\delta_0$, we cannot use PN approximations because it violates the conditions that the value of 1-PN part should be much less than 0-PN part.
%Note that signals from such large galaxies is difficult to be detected because the orbital time period is very high. 
The important results we get here are,
\begin{itemize}
\item The average energy radiation rate that we have derived is applicable for every EMRI which is surrounded by a matter distribution that can give power law potential in the region around the supermassive compact object. We show that the average energy radiation rate is higher in general PL potential case as compared to that in KN potential.
\item We use the comparison of orbital frequency of an EMRI in PL and KN potentials to show how matter distribution (e.g. dark matter distribution) causes change in the signals. This can be used to further investigate the quantities like frequencies of the gravitational waves radiated from such an EMRI. The deviation of the observed orbital frequencies from the frequency of a similar binary in KN potential can be used to evaluate the value of $\delta$ and that can be used to find out the effective potential and subsequently, can calculate the mass density profile around the binary.
\end{itemize}

\section{Acknowledgement}
C.N.G. would like to acknowledge the support of the International Center for Cosmology, CHARUSAT, India for funding the work.
%\bibliographystyle{plain}


%
% ****** End of file apssamp.tex ******

\end{document}